\documentclass[%
 floatfix,
 reprint,
superscriptaddress,
nobibnotes,
 amsmath,amssymb,
 aps,
prb,
]{revtex4-2}

\usepackage{url}

\usepackage{graphicx}
\usepackage{subcaption}
\usepackage{hyperref}
\usepackage[english]{babel}

\draft 

\begin{document}


\title{A Cryogenic Geophone with stable response from Room Temperature to millikelvin} 


\def\andname{}
\author{Stefano Pirro}
\affiliation{INFN Laboratori Nazionali del Gran Sasso, Via G. Acitelli, 22, Assergi (AQ), 67100, Italy}

\author{Marco Ricci}
\email{marco.ricci@roma1.infn.it}
\affiliation{INFN Sezione di Roma, P.le A. Moro 2, Roma, 00185, Italy}

\author{Simone Cioccolini}
\affiliation{INFN Laboratori Nazionali del Gran Sasso, Via G. Acitelli, 22, Assergi (AQ), 67100, Italy}

    


\begin{abstract}
With the development of pulse-tube cryocoolers, the availability and use of low-temperature environments have increased exponentially in recent years. However, replacing liquid helium with a mechanical cooling device has, since its introduction, highlighted the need to attenuate the mechanical vibrations generated by the cryocooler itself.
In the case of dilution refrigerators, which are capable of reaching temperatures on the order of few  mK, the issue of mechanical vibrations reaching the experimental setup is, in many situations, extremely critical. This has led to the development of several customized experimental approaches aimed at attenuating vibrations. However, the inertial measurement of vibrations at cryogenic and, in particular, ultracryogenic temperatures has relied on customized sensors, since no commercially available, calibrated instruments exist for measurements at these temperatures.
Nevertheless, many research groups employ commercial geophones that are recalibrated at low temperatures. This approach, however, is not free from systematic errors and presents significant challenges in quantitatively assessing the absolute response.
In this work, we describe the development of cryogenic geophones exhibiting an almost temperature-independent response, from room temperature down to approximately $10$ mK.
\end{abstract}

\pacs{}
\maketitle

\section{Introduction}

\label{sec:Introduction}

The widespread adoption of dry refrigerators and Pulse-Tube (PT) cryocoolers has revolutionized experimental low-temperature physics by eliminating the operational reliance on liquid helium \cite{DRY_Dilution}. 
As early as the end of the last century, it became evident that vibrations generated by the circulation of helium gas in liquid-helium-based dilution refrigerators could produce acoustic noise, which is subsequently converted into thermal noise in bolometric
detectors \cite{PIRRO_Vibration}. This effect motivated the development of dedicated mechanical decoupling systems capable of operating at millikelvin temperatures, with the aim of suppressing vibration-induced detector noise and improving energy resolution \cite{PIRRO_Vibration2}. 
About a decade later, the advent of pulse-tube-based dry dilution refrigerators brought renewed attention to vibration mitigation \cite{Atomic_Resolution, OLIVIERI201773, Lee2018VibrationMF}, as the rotary valve and cyclic compression of the working gas were found to generate substantially larger mechanical vibrations and microphonic noise than those encountered previously in conventional liquid-helium systems

Mechanical dissipation induced by vibration directly converts into parasitic heat loads, degrading the base temperature and spoiling the micro-vibrational environment. This issue is particularly acute across a wide spectrum of low-temperature applications: from milliKelvin environments, including rare-event searches (e.g., bolometric dark matter \cite{COSINUS_Vibration} and neutrinoless double-beta decay experiments \cite{EDELWEISS_Vibration, EDELWEISS_Vibration2c, NUCLEUS_Vibration, CUORE_PT}) and sub-Kelvin quantum computing platforms \cite{Mechanical_induced_Vibrations_Quibit}, to cryogenic systems operating in the $\approx 10$K regime, such as underground interferometric gravitational wave detectors \cite{Yamada_HLVIS}. Consequently, designing effective passive or active isolation schemes requires absolute, low-frequency, and temperature-independent inertial vibration measurement directly at cryogenic and ultracryogenic stages.

Commercially available vibration sensors, such as inertial piezoelectric accelerometers, are typically operable down to liquid nitrogen temperatures (77 K) \cite{ICP, Deng_Design}, and some are certified for operation at liquid helium temperatures (4.2 K) \cite{Endevco, Kistler}. However, these cryogenic accelerometers have poor sensitivity below 10 Hz. Moreover, models designed to operate below 70 K show higher noise levels, largely because the charge amplifier is not integrated into the device itself. Consequently, the length and mechanical vibrations of the cable connecting the accelerometer to the preamplifier significantly degrade both the sensitivity and the overall measurement quality.

Geophones, widely used around the world, offer an alternative approach for inertial vibration measurement at low temperatures. A commercial (vertical)  geophone consists of a permanent magnet rigidly attached to a supporting mechanical structure and  2 coils (in a Helmholtz configuration)  of wire wound around a cylindrical structure \cite{Sheriff_Geldart_1995}. The coil and the cylindrical structure represent the so-called inertial mass (IM) that is suspended by spiral springs such that it is free to oscillate relative to the magnet under the action of gravitational and restoring elastic forces, which act in opposite directions. The relative motion between the magnet and the coil induces an electromotive force in the coil, from which the characteristics of the originating vibration can be inferred.

Commercial geophones are specified by manufacturers for an operating temperature range typically between $-55$ °C and $+85$ °C.

In the last two decades, however, a large variety of measurements were reported operating commercial geophones at low and ultralow temperatures \cite{Blair_1991, SCHMORANZER_vibrations, Ye_Vibrations}.
In general, the results reported in these publications exhibit several critical aspects and, in some cases, inconsistencies. None of them, however, discusses or explains the operation of geophones at low temperatures, nor the fundamental reasons why their characteristics change, and, in the case of low-frequency geophones, why they may even cease to operate. Likewise, no work proposes or suggests methods to improve or adapt geophones for cryogenic use in a way that is independent of temperature-dependent recalibration or active feedback.

Instead, these studies mainly describe how to adapt a commercial device, by means of software-based signal analysis or specific calibration algorithms applied on a case-by-case basis, to perform measurements beyond the manufacturer’s specified operating range \cite{KOHL_extending, Feedback}.

The main conclusions that can be drawn from these publications, and which they share in common, are as follows:

\begin{itemize}
    \item Commercial geophones that “operate” at cryogenic temperatures exhibit, when cooled, a natural frequency (and thus sensitivity) not lower than about 5 Hz.
    \item The response of the geophone is temperature-dependent and must be recalibrated each time using additional electronic instrumentation.
     \end{itemize}

\begin{figure*}
    \centering
    \includegraphics[width=0.75\textwidth]{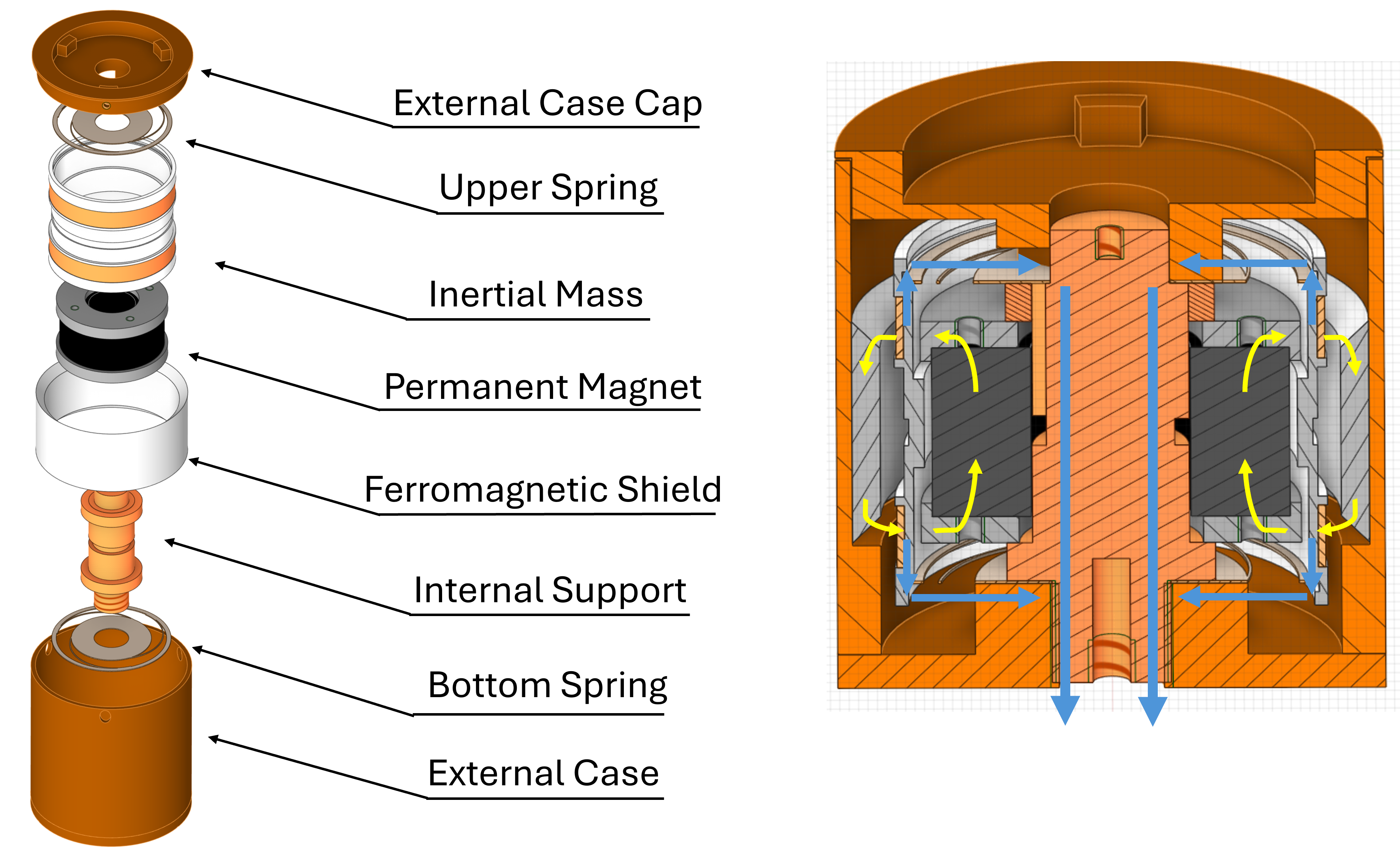}
    \caption{Structural layout of the modified cryogenic geophone. Left: exploded view showing the copper chassis, Macor$^{\circledR}$ inertial mass, coils, copper internal support, and planar springs. Right: cross-sectional view illustrating the magnetic circuit (yellow arrows) and the internal conductive thermalization paths (blue arrows).}
    \label{fig:CH7_GeophoneComp}
\end{figure*}

Therefore, the geophones developed so far do not provide reliable performance as the temperature decreases, and the underlying reasons why, in many cases, they cease to work properly remain unclear.

To solve these limitations, this work presents a dedicated structural modification strategy for standard low-frequency geophones, tailored to operate reliably down to cryogenic temperatures. 

The paper is organized as follows: Section~\ref{sec:design_innovations} provides a detailed analysis of the physical mechanisms governing cryogenic degradation and describes the structural solutions implemented to achieve a temperature-independent response. Section~\ref{sec:calibration} and Section~\ref{sec:results} present the electromechanical calibration and performance results from room temperature down to ultracryogenic levels, comparing the custom \textit{CryoGeo} prototype with a standard commercial unit to demonstrate its superior thermal stability. Section~\ref{sec:noise} covers the experimental noise analysis, highlighting the low-frequency resolution achievable down to millikelvin temperatures. Finally, Section~\ref{sec:conclusions} summarizes the main findings, highlights the patenting and TRL status, and outlines potential application scenarios.

\section{Working Principle and Cryogenic Modifications}
\label{sec:design_innovations}

The internal mechanical architecture of the modified geophone is illustrated in Fig.~\ref{fig:CH7_GeophoneComp}, showing both an exploded view and a cross-sectional view of the assembly. The sensor consists of a central internal support rigidly fastened to an outer cylindrical chassis \cite{Sheriff_Geldart_1995}. A permanent magnet is mounted onto the central support, enclosed between two ferromagnetic end caps and surrounded by an outer ferromagnetic shield to complete the magnetic circuit, creating a uniform magnetic field within the annular gap.

The inertial mass, which carries the sensing coils, is suspended from the central support by two planar spiral springs (upper and lower). Under standard room-temperature conditions, the sensor behaves as a damped single-degree-of-freedom mechanical harmonic oscillator. Its complex velocity transfer function $H(\omega) = V(\omega)/v(\omega)$, describing the voltage readout per unit input velocity \cite{Hons}, is given by:
\begin{equation}
\label{eq:Transfer_function}
    H(\omega) = G \frac{\left( \frac{\omega}{\omega_0} \right)^2}{1 - \left( \frac{\omega}{\omega_0} \right)^2 + i\frac{1}{Q}\left(\frac{\omega}{\omega_0}\right)},
\end{equation}
where $G = Bl$ is the open-circuit transduction gain ($B$ being the magnetic field strength and $l$ the effective coil length immersed in the gap), $\omega_0 = 2\pi f_0 = \sqrt{k/m}$ is the natural angular frequency defined by the spring stiffness $k$ and suspended mass $m$, and $Q$ is the quality factor. 

In experimental measurements, the sensor's magnitude response defines the frequency-dependent sensitivity $S(\omega)$, measured in $\text{V}/(\text{m}/\text{s})$, which corresponds directly to the absolute value of the transfer function $ S(\omega) = |H(\omega)| $.

While conventional geophones are optimized for room-temperature operation (typically targeted to a critically damped condition $Q \approx 1/\sqrt{2}$), cooling the device to cryogenic temperatures introduces three major failure mechanisms. To overcome these limitations, specific structural and material modifications were implemented.

\subsection{Eddy-Current Overdamping and Mass Insulation}

In standard commercial geophones, the inertial mass body is machined from conductive metals (such as aluminum or brass). As the mass oscillates within the magnetic gap, the relative motion induces eddy currents with a current density $\vec{J}_{\text{eddy}} = (\vec{v}\times\vec{B})/\rho$, generating an opposing magnetic force density $\vec{f} = \vec{J}_{\text{eddy}}\times\vec{B}$ \cite{Huan_Prediction}. At low temperatures, the metallic electrical resistivity $\rho$ drops drastically, causing a severe increase in eddy-current damping that reduces $Q$.

\begin{figure}
    \centering
    \includegraphics[width=5cm, angle=-90, clip, trim=2cm 0.3cm 1.8cm 0.5cm]{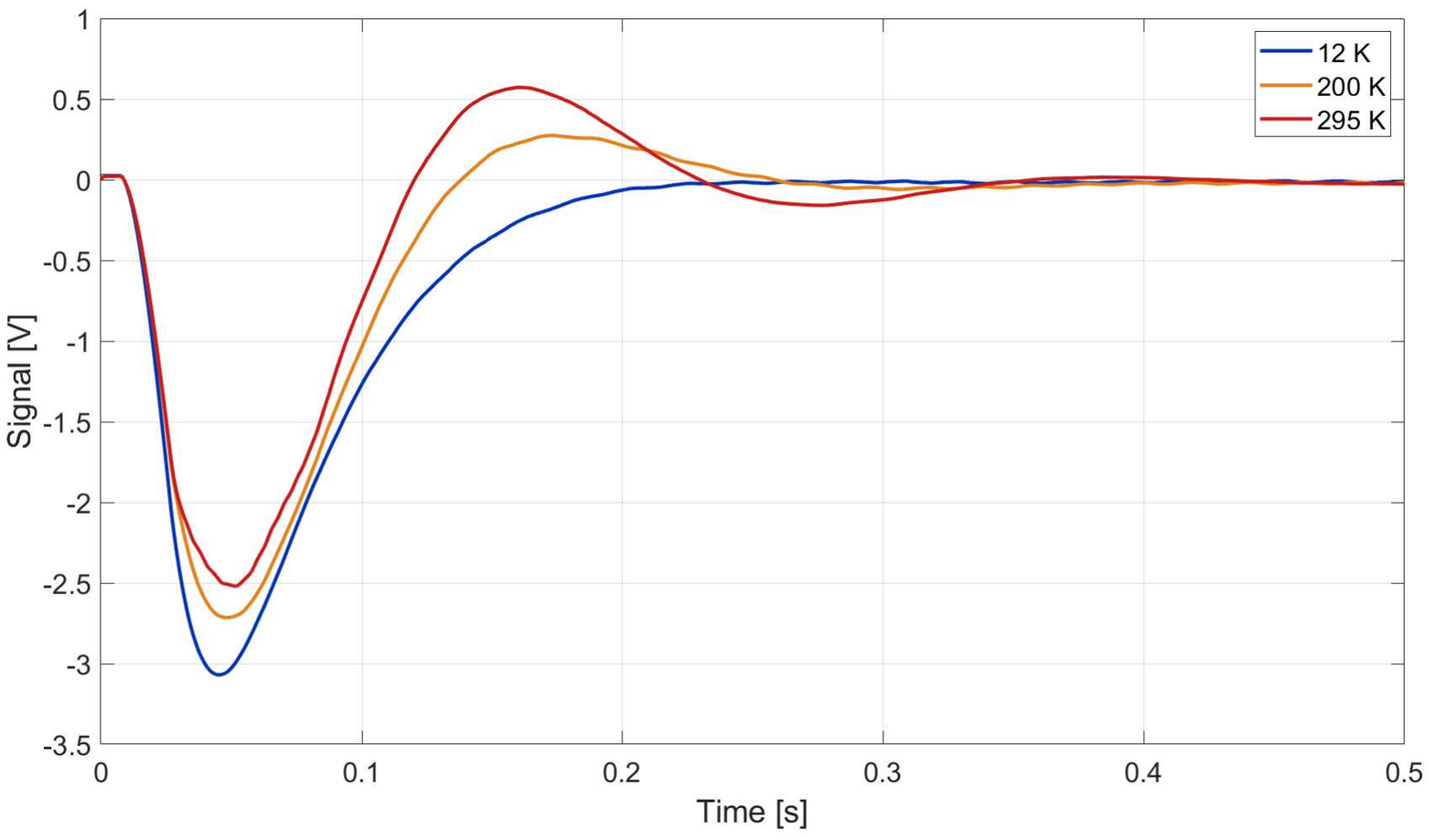}
    \caption{Transient voltage pulse response of a commercial $4.5\text{~Hz}$ geophone \cite{GeoSpace} subjected to an identical electrical step excitation at $295\text{~K}$ (red), $200\text{~K}$ (orange), and $12\text{~K}$ (blue). The progressive reduction of oscillations directly illustrates the severe eddy-current overdamping caused by the drop in metallic resistivity at cryogenic temperatures.}
    \label{fig:Geo_Space_4_5_Hz}
\end{figure}

This progressive temperature-dependent overdamping is clearly demonstrated in Fig.~\ref{fig:Geo_Space_4_5_Hz}, which shows the electrical transient step response of a commercial $4.5\text{~Hz}$ geophone \cite{GeoSpace} measured at $295\text{~K}$, $200\text{~K}$, and $12\text{~K}$. As the temperature drops, its oscillation is progressively suppressed: already at $200\text{~K}$ the transient shows a noticeable decay in oscillation amplitude, culminating at $12\text{~K}$ in a heavily overdamped, non-oscillatory response.

To eliminate this parasitic electromagnetic damping, the metallic inertial mass body was replaced with Macor$^{\circledR}$ (Corning Inc., USA), a machinable glass-ceramic that is non-conductive, vacuum-compatible, and possesses a mass density ($\rho = 2.52\text{~g/cm}^3$) closely matching that of aluminum. This design choice preserves the original mechanical dimensions and suspended mass while completely suppressing eddy-current dissipation across the entire cryogenic range.

However, eliminating eddy currents leaves the oscillator dissipation dominated almost entirely by the intrinsic internal friction of the mechanical suspension. This results in a very high quality factor ($Q \gg 1$), causing prolonged transient ring-down oscillations that limit the sensor's temporal resolution for closely spaced vibrational transients.

To overcome this, the desired quality factor can be tailored independently using a parallel shunt resistor $R_S$ connected across the output terminals. In this shunted configuration, the transfer function parameters defined in Eq.~\ref{eq:Transfer_function} are modified as follows \cite{Amirikas_CryoAbsolute}:
\begin{equation}
\label{eq:loaded_parameters}
\begin{aligned}
    G' &= \frac{R_S}{R_S + R_C} G, \\[6pt]
    Q' &= \frac{Q}{1 + \frac{G^2 Q}{2\pi f_0 m (R_C + R_S)}},
\end{aligned}
\end{equation}
where $R_C$ is the internal resistance of the sensing coil and $m$ is the suspended mass.

While this shunting strategy provides flexible tuning of the quality factor (e.g., toward the critically damped target $Q' \approx 0.707$), it introduces a trade-off: the effective transduction gain $G'$ is reduced by the resistive current-divider ratio. 
Furthermore, as the temperature decreases, the coil resistance $R_C(T)$ decreases rapidly following the Residual Resistivity Ratio (RRR) of copper \cite{Calatroni_RRRCopper}. According to Eq.~\ref{eq:loaded_parameters}, this drop in $R_C$ increases the electromagnetic damping term $\propto (R_C + R_S)^{-1}$, leading to a temperature-dependent shift in $Q'$ that can, however, be precisely compensated if needed by adjusting the external shunt resistance $R_S$.

\subsection{Elastic Modulus Drift and Mass Displacement Dynamics}

As the temperature decreases, the Young's modulus $E(T)$ of the spring material (e.g., Beryllium Copper) increases \cite{CuBe_Young}, raising the spring stiffness $k(T)$. For vertical geophones, pre-tensioned to hold the mass in equilibrium under gravity at room temperature ($k_0$), cooling to temperature $T$ induces higher spring restoring forces that pull the mass upward by a displacement $d(T)$:
\begin{equation}
\label{eq:displacement}
\begin{aligned}
    d(T) &= \frac{g}{4\pi^2} \left( \frac{k(T) - k_0}{k(T)} \right) \frac{1}{f_0^2} \\
         &\approx 248 \cdot \left(\frac{\Delta k}{k(T)}\right) \frac{1}{f_0^2} \quad [\text{mm}].
\end{aligned}
\end{equation}
Because this vertical displacement scales inversely with $f_0^2$, low-frequency geophones ($f_0 \le 4.5\text{~Hz}$) undergo severe upward mechanical shifts that easily exceed the available internal clearance ($\sim \pm \>1\text{--}3\text{~mm}$). 

\begin{figure}
    \centering
    \includegraphics[width=5cm, angle = -90,clip, trim = 3.5cm 2.4cm 3cm 1.4cm]{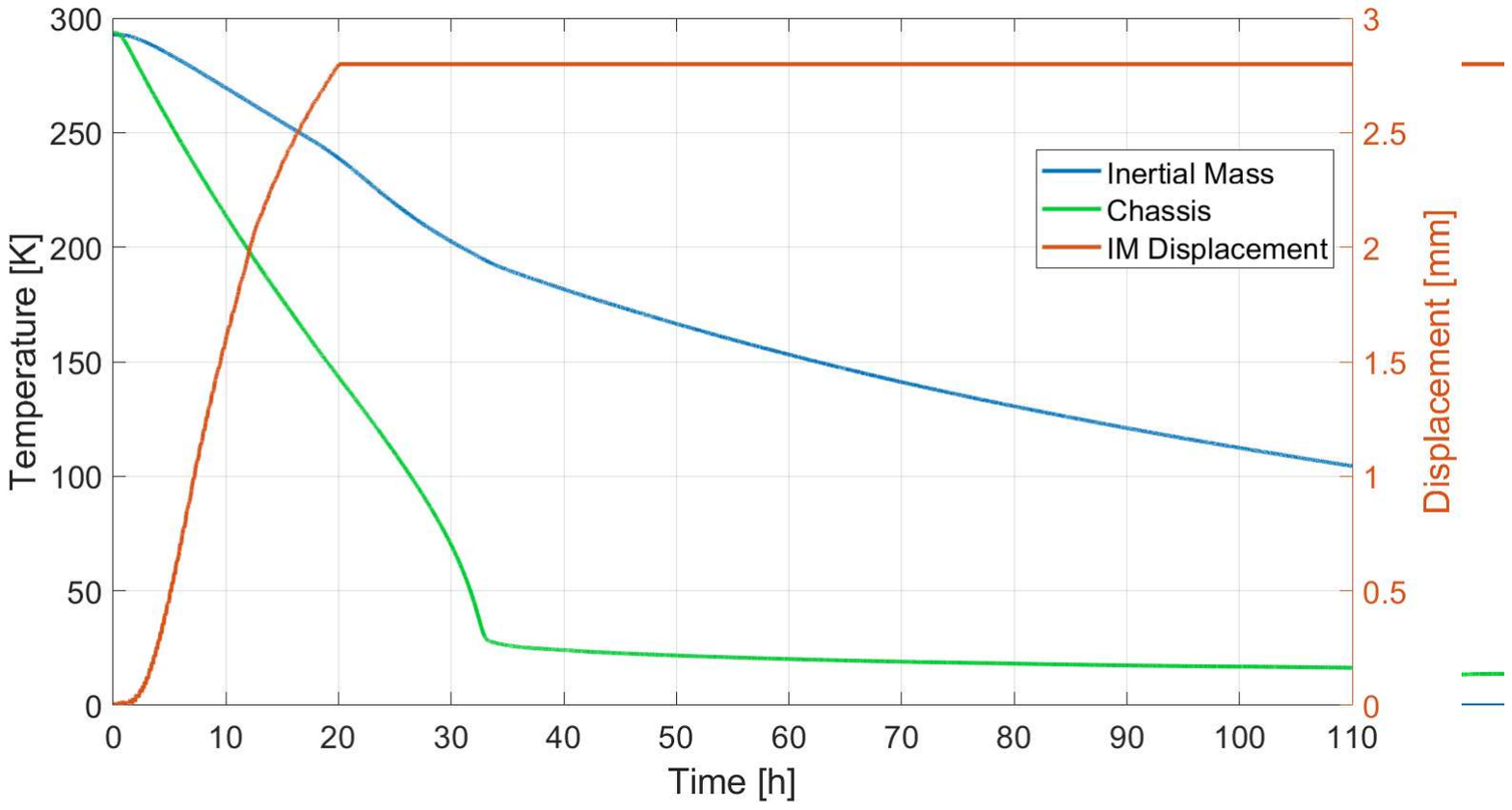}
    \caption{Cooling dynamics and mechanical displacement of a commercial $1\text{ Hz}$ geophone \cite{Sercel} mounted on a Pulse Tube cold plate. At $T_{IM} \approx 240\text{ K}$, the mass displacement reaches the $2.8\text{ mm}$ limit, causing mechanical contact.}
    \label{fig:Mass-displacement}
\end{figure}

Fig. ~\ref{fig:Mass-displacement} illustrates this failure during the cooldown of a standard $1\text{~Hz}$ commercial geophone (Sercel LC-4 \cite{Sercel}). As the inertial mass cools below $240\text{~K}$, it moves upward by $2.8\text{~mm}$ until it makes contact with the upper mechanical end-stop, completely freezing the oscillator motion.

\begin{figure}
    \centering
    \includegraphics[width= 5cm, angle=-90,clip, trim = 3cm 2.4cm 3cm 1.4cm]{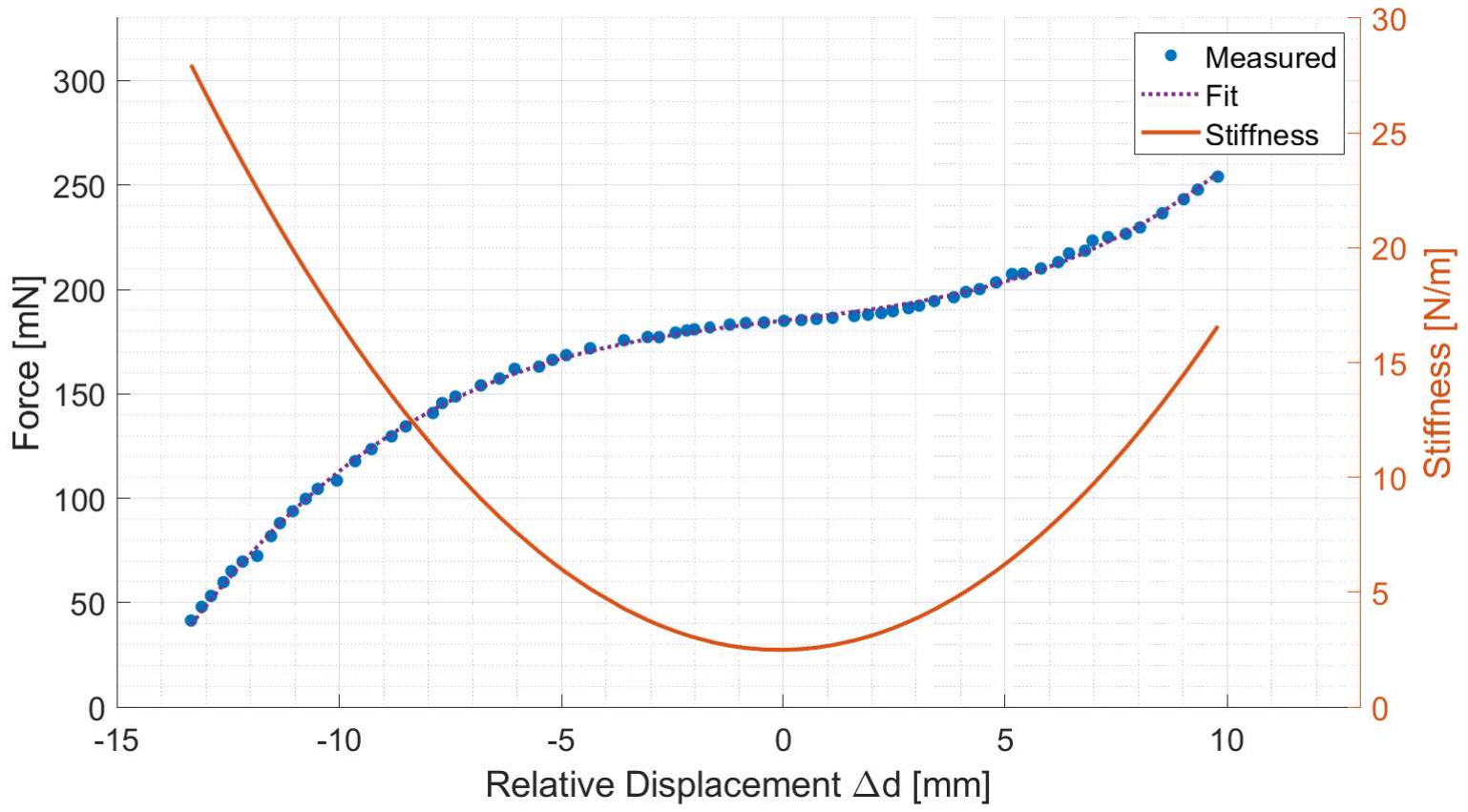}
    \caption{Measured force-displacement behavior of a geophone suspension spring (scatter points) fitted with a third-order polynomial (dotted line). The solid curve represents the local derivative stiffness (right axis).}
    \label{fig:Stiffness}
\end{figure}

Moreover, this thermal shift in equilibrium operational position results in a non-linear distortion of the resonant frequency $f_0(T)$. 
As a reference, Fig.~\ref{fig:Stiffness} presents the measured force-displacement characteristic of one of the planar spiral springs used in our new \textit{CryoGeo} assembly. The experimental data were fitted with a third-order polynomial, whose local derivative yields the displacement-dependent stiffness $k(d) = k_0 + 2\alpha d + 3\beta d^2$.

By incorporating the temperature dependence of Young's modulus, the effective spring stiffness at temperature $T$, denoted as $k_T(d(T))$, explicitly exhibits a coupled elastic and kinematic temperature dependence:
\begin{equation}
\label{eq:coupled_stiffness}
    k_T(d(T)) = \frac{E(T)}{E_0} \cdot \left( k_0 + 2\alpha \, d(T) + 3\beta \, d(T)^2 \right),
\end{equation}
where $k_0$ is the nominal linear spring stiffness coefficient at $300\text{~K}$ around the origin ($d = 0$), $\alpha$ and $\beta$ are higher-order non-linear geometric elastic coefficients, and $d(T)$ is the temperature-driven operational position along the vertical axis.

In an uncompensated geophone, these two distinct phenomena act synergistically to cause a severe coupled non-linear increase in $f_0(T)$:
\begin{enumerate}
    \item \textbf{Elastic Modulus Scaling:} The term  $E(T)/E_0$ increases directly with decreasing temperature, shifting the entire stiffness curve upward.
    \item \textbf{Non-Linear Kinematic Drift:} Simultaneously, the upward movement $d(T)$ displaces the equilibrium position away from the central flat region ($d \approx 0$), forcing operation deeper into the non-linear branches where $\big(2\alpha d(T) + 3\beta d(T)^2\big) > 0$ on the stiffness curve (Fig.~\ref{fig:Stiffness}).
\end{enumerate}

To maintain full mechanical travel and stabilize the resonant frequency across the entire operational temperature range, two complementary solutions were implemented:
\begin{enumerate}
    \item The internal downward mechanical clearance was expanded to accommodate greater physical clearance without mechanical contact against the housing.   
    \item A dedicated thermal-compensation mass increment $\Delta m \approx m (\Delta E / E)$ was directly integrated into the design of the inertial mass assembly. The target ($m_{\text{tot}} = m_0 + \Delta m$) was precisely achieved by tailoring the internal geometry of the Macor$^{\circledR}$ body and coil weight. At $300\text{~K}$, this intentional additional load sets the equilibrium position at a negative offset ($d_0 < 0$). Upon cooling, as the increased spring stiffness pulls the mass upward ($d(T) \to 0$), the operational position moves toward the minimum of the local stiffness curve where $\big(2\alpha d(T) + 3\beta d(T)^2\big) \to 0$. This geometric reduction in local non-linear stiffness dynamically compensates for the global rise in $E(T)$, stabilizing the resonant frequency across the entire temperature range from $10\text{~mK}$ to $300\text{~K}$:
    
\begin{equation}
\label{eq:stabilized_f0}
    f_0(T) = \frac{1}{2\pi} \sqrt{\frac{k_T(d(T))}{m + \Delta m}} \approx \text{const.}
\end{equation}
\end{enumerate}

\subsection{Thermalization at Low Temperatures}
In commercial geophones, to have a robust and straightforward assembly procedure, the inertial mass is electrically and, therefore, thermally isolated from the outer housing. As shown by the thermal curves in Fig. \ref{fig:Mass-displacement}, this implies the temperature of the mass drops much more slowly than that of the chassis, cooling almost exclusively via thermal radiation at low temperatures. In milliKelvin environments (such as dilution refrigerators), an un-thermalized mass acts as a persistent thermal leak ($\sim \mu\text{W}$) over extended periods.

To ensure efficient conductive thermalization, both the outer chassis and the central internal support were fabricated from high-purity copper (Fig. \ref{fig:CH7_GeophoneComp}). Heat is extracted from the Macor$^{\circledR}$ mass directly through conductive thermal paths provided by the beryllium-copper suspension springs connected directly to the central copper support.

\section{Calibration Methodologies}
\label{sec:calibration}

A geophone is fully calibrated when the three fundamental parameters defining its transfer function (the transduction gain $G$, the natural frequency $f_0$, and the quality factor $Q$) are experimentally determined \cite{Rodgers_Step}. While several calibration techniques have been established in the literature, their feasibility varies significantly when applied in cryogenic and sub-Kelvin environments.

The most direct approach relies on exciting the sensor with a mechanical shaker supplying a controlled input motion across a frequency sweep, while simultaneously measuring the generated output voltage \cite{Shaker_calib, shaker_masahiro}. Although this method directly implements the physical definition of the transfer function, its execution becomes challenging, and often impractical, in cryogenic systems due to spatial constraints, mechanical thermal bridges, and complex cryogenic setup requirements. While applicable at room temperature or in dedicated facilities, shaker-based calibration is not scalable for routine \textit{in-situ} calibration.

Van Kann and Winterflood \cite{CalibAssoluta, Amirikas_CryoAbsolute} developed an absolute, reciprocity-based calibration technique that measures the voltage-divider ratio of a circuit formed by the geophone in series with a known reference resistor. By driving the system with broadband white noise over the frequency band of interest, the measured response yields the three transfer function parameters along with the coil resistance and inductance. Despite its rigor, this method requires a multi-parameter non-linear fit with five free parameters, increasing fit-related uncertainties. Furthermore, achieving an adequate Signal-to-Noise Ratio (SNR) at very low frequencies ($0.1\text{--}1\text{~Hz}$) demands long integration times with continuous white-noise excitation, introducing non-negligible thermal dissipation: a critical constraint in milliKelvin platforms.

To overcome these operational constraints, the step-current (or pulse-response) method \cite{Rodgers_Step, BowdenPulse, CalibrationL4CPulse} emerges as the most suitable technique for cryogenic \textit{in-situ} characterization. In this approach, a small DC current $I_{\text{ext}}$ is injected into the geophone coils to produce an initial static displacement of the inertial mass. A fast Double-Pole Double-Throw (DPDT) switch then disconnects the excitation circuit and connects the sensor directly to the data acquisition system. The resulting free transient ring-down is recorded (as shown in Fig. \ref{fig:Geo_Space_4_5_Hz}) and fitted using the standard damped harmonic oscillator equation to extract $f_0$ and $Q$. The absolute transduction constant $G$ is then derived via the relation:
\begin{equation}
    \frac{G^2}{m} = \frac{A}{I_{\text{ext}}} \, \omega_0 \sqrt{1 - \left( \frac{1}{2Q} \right)^2},
    \label{eq:step_current}
\end{equation}
where $A$ is the fitted initial voltage amplitude of the transient decay.

Compared to other methods, the step-current technique offers significant advantages for low-temperature applications:
\begin{itemize}
    \item \textbf{Minimal Instrumentation \& Heat Load:} It requires no dedicated internal actuators or mechanical shakers, relying solely on a simple DC supply and a low-dissipation pulse drive.
    \item \textbf{High Computational Robustness:} The transient response is fitted with only three free parameters ($\omega_0, Q, A$), making the data extraction fast, reliable, and immune to high-order fit instabilities.
    \item \textbf{In-Situ Compatibility:} It can be executed rapidly during cooldown cycles to track temperature-dependent parameter drifts without disturbing the cryogenic thermal equilibrium.
\end{itemize}

It should be noted that both electrical calibration approaches (the reciprocity technique and the step-current method) exhibit a fundamental parameter degeneracy, as the electrical response scales directly with the ratio $G^2/m$. Inferring the absolute transduction gain $G$ requires independent knowledge of the inertial mass $m$: it is temperature-independent and can be accurately measured \textit{a priori} during sensor assembly. Consequently, the step-current method was adopted as the primary technique for the cryogenic characterization presented in this work.
\section{Experimental Results and Discussion}
\label{sec:results}

\begin{figure*}
    \centering
    \hspace*{-3.4cm}
    \begin{subfigure}[b]{0.330\textwidth}
        \centering
        \includegraphics[width=0.9\textwidth, angle=-90, clip, trim=2cm 0.4cm 1.6cm 0.85cm]{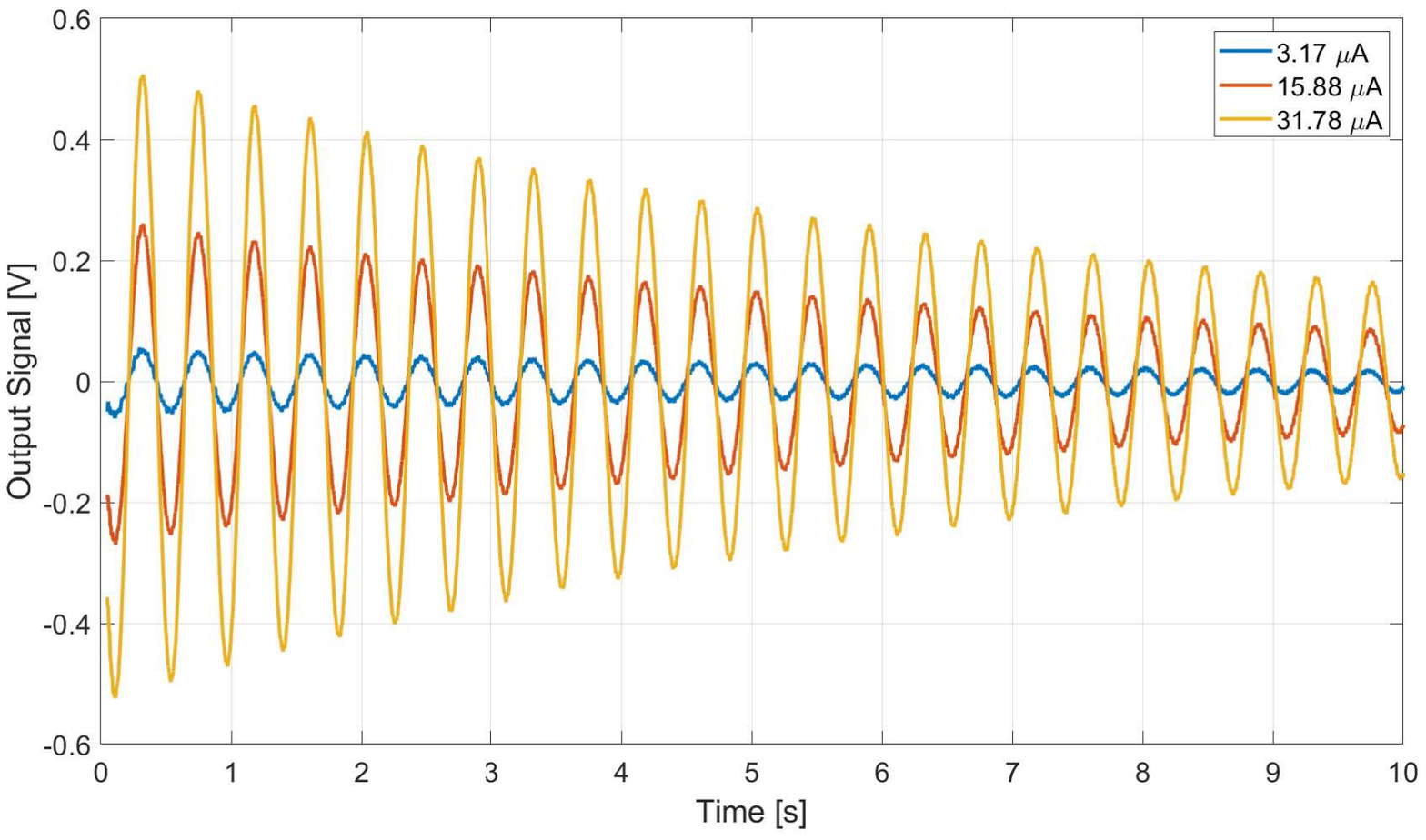}
        \caption{}
        \label{fig:step_responses}
    \end{subfigure}%
    \hspace{0.17\textwidth}%
    \begin{subfigure}[b]{0.33\textwidth}
        \centering
        \includegraphics[width=5.3cm, height=9.2cm, angle=-90, clip, trim=2cm 0cm 1.2cm 1cm]{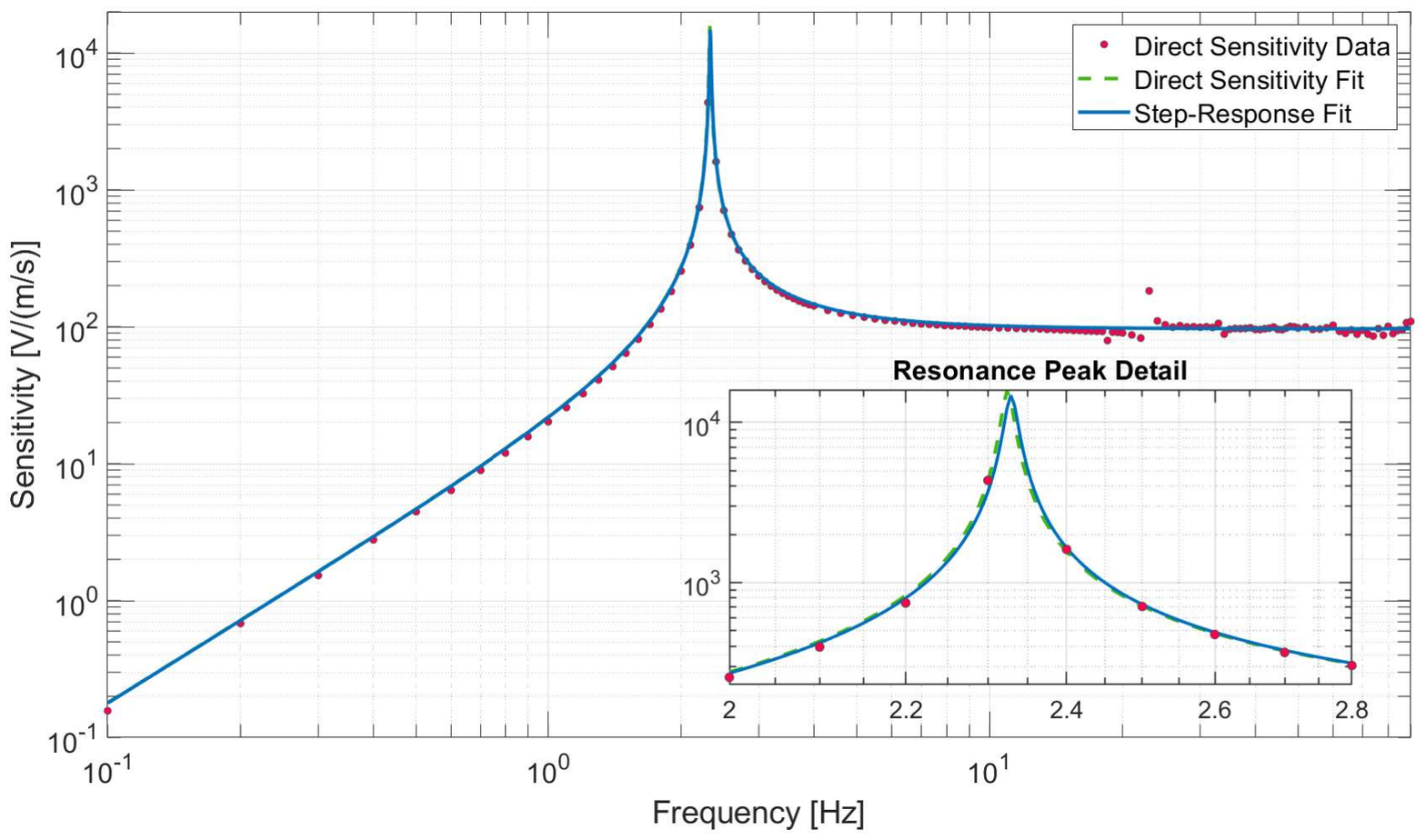}
        \caption{}
        \label{fig:tf_comparison}
    \end{subfigure}
    \caption{Room-temperature ($T = 295\text{ K}$) cross-validation of the step-current method: (a) Step-current transient ring-down responses for different excitation current amplitudes ($3.17\mu$A, $15.88\mu$A, and $31.78\mu$A), demonstrating strictly linear oscillator dynamics. (b) Sensitivity comparison between direct frequency-sweep shaker measurements (red markers), the corresponding analytical fit (green dashed curve), and the prediction reconstructed from step-current transient parameters (blue solid curve). The inset highlights the sharp resonance peak at $f_0 \approx 2.33\text{ Hz}$.}
    \label{fig:room_temp_validation_composite}
\end{figure*}

To validate the structural modifications and verify the operational stability of the custom cryogenic geophone across the target temperature range, a long experimental campaign was conducted using two complementary calibration techniques: direct mechanical excitation via a shaker system and the step-current method.

To provide a clear comparative evaluation, the characterization analysis focuses on three key representative configurations:
\begin{enumerate}
    \item \textbf{Cryogenic Geophone (\textit{CryoGeo}):} The modified sensor equipped with a Macor$^{\circledR}$ inertial mass, enlarged internal clearance, copper structure, and a compensated mass.
    \item \textbf{Shunted Cryogenic Geophone (\textit{CryoGeo Shunted}):} The previous sensor, but read out in parallel with a room-temperature (shunt) resistor $R_S = 100~\text{k}\Omega$ (constant across all temperatures investigated) to reduce the quality factor.
    \item \textbf{Commercial Benchmark (\textit{GeoSpace}):} A standard commercial GeoSpace GS-ONE LF $4.5$Hz geophone \cite{GeoSpace} used as a control baseline.
\end{enumerate}

\subsection{Room-Temperature Baseline and Method Cross-Validation}
\label{sec:room_temp_validation}

A room-temperature baseline characterization was initially performed at $T = 295\text{~K}$ to cross-validate the step-current method against the direct shaker excitation.\\
For the direct approach, the sensors were mounted on an electrodynamic shaker controlled by custom automated LabVIEW (National Instruments\textsuperscript{TM}) software, which controlled a sinusoidal frequency sweep to drive the shaker platform. During the sweep, the system simultaneously recorded the voltage output of the geophones and the mechanical excitation of the shaker plate. The absolute base displacement, identical to the motion of the rigidly coupled geophone chassis, was independently monitored using a high-precision fiber-optic displacement sensor (Philtec Inc. \cite{fiberoptic}). The optical fiber was calibrated with a dedicated procedure by using a calibrated height gauge with fine adjustment.  \\
For the step-current method, DC excitation currents ranging from $3.17~\mu\text{A}$ to $79.5~\mu\text{A}$ were applied and abruptly cut off to excite the transient ring-down motion.

As a reference, Fig. \ref{fig:step_responses} shows the synchronized transient ring-down responses recorded for three different excitation currents ($3.17\mu$A, $15.88\mu$A, and $31.78\mu$A) for the cryogenic geophone. The perfect scaling of amplitude without distortion confirms the linearity of the mechanical suspension across the tested range.

As summarized in Table \ref{tab:room_temp_comp}, the parameters extracted from both techniques show excellent agreement within experimental uncertainties. Fig. \ref{fig:tf_comparison} depicts a direct comparison between the transfer function measured via shaker sweep and the theoretical transfer function reconstructed from the step-current ring-down parameters (following Eq. \ref{eq:Transfer_function}). The inset demonstrates that both methods align remarkably well at the resonance peak ($f_0 \approx 2.33$Hz). 

Notably, the step-current method yields significantly smaller uncertainties on the quality factor $Q$ for high-$Q$ configurations (e.g., Unshunted CryoGeo), where the limited frequency point density near the sharp resonance peak during shaker sweeps introduces higher fit variance. This confirms the step-current approach as a highly accurate, non-invasive, and computationally robust tool for sensor calibration.

\begin{table*}
\caption{Comparison of sensitivity parameters ($G$, $f_0$, $Q$) obtained at $295$K via Direct Shaker Excitation and Step-Current Method.}
\label{tab:room_temp_comp}
\centering
\begin{tabular}{l ccc}
\hline
\hline
\textbf{Sensor Configuration} & \textbf{$G$ [V/(m/s)]} & \textbf{$f_0$ [Hz]} & \textbf{$Q$ [-]} \\
\hline
\textit{Direct Shaker Method} & & & \\
CryoGeo (Unshunted) & $96.7 \pm 1.0$ & $2.32 \pm 0.05$ & $164 \pm 36$ \\
CryoGeo (Shunted)   & $74.9 \pm 0.9$ & $2.34 \pm 0.01$ & $6.69 \pm 0.15$ \\
GeoSpace $4.5\text{ Hz}$      & $98.3 \pm 0.5$ & $4.63 \pm 0.04$ & $1.09 \pm 0.01$ \\
\hline
\textit{Step-Current Method} & & & \\
CryoGeo (Unshunted) & $96.6 \pm 0.6$ & $2.33 \pm 0.01$ & $153.3 \pm 1.2$ \\
CryoGeo (Shunted)   & $74.7 \pm 0.6$ & $2.33 \pm 0.01$ & $6.73 \pm 0.01$ \\
GeoSpace $4.5\text{ Hz}$      & $94.8 \pm 0.7$ & $4.72 \pm 0.04$ & $1.20 \pm 0.01$ \\
\hline
\hline
\end{tabular}
\end{table*}

\subsection{Cryogenic Experimental Setup and Calibration Methodology}
\label{sec:cryo_setup}

\begin{figure}
    \centering
    \includegraphics[width=\columnwidth]{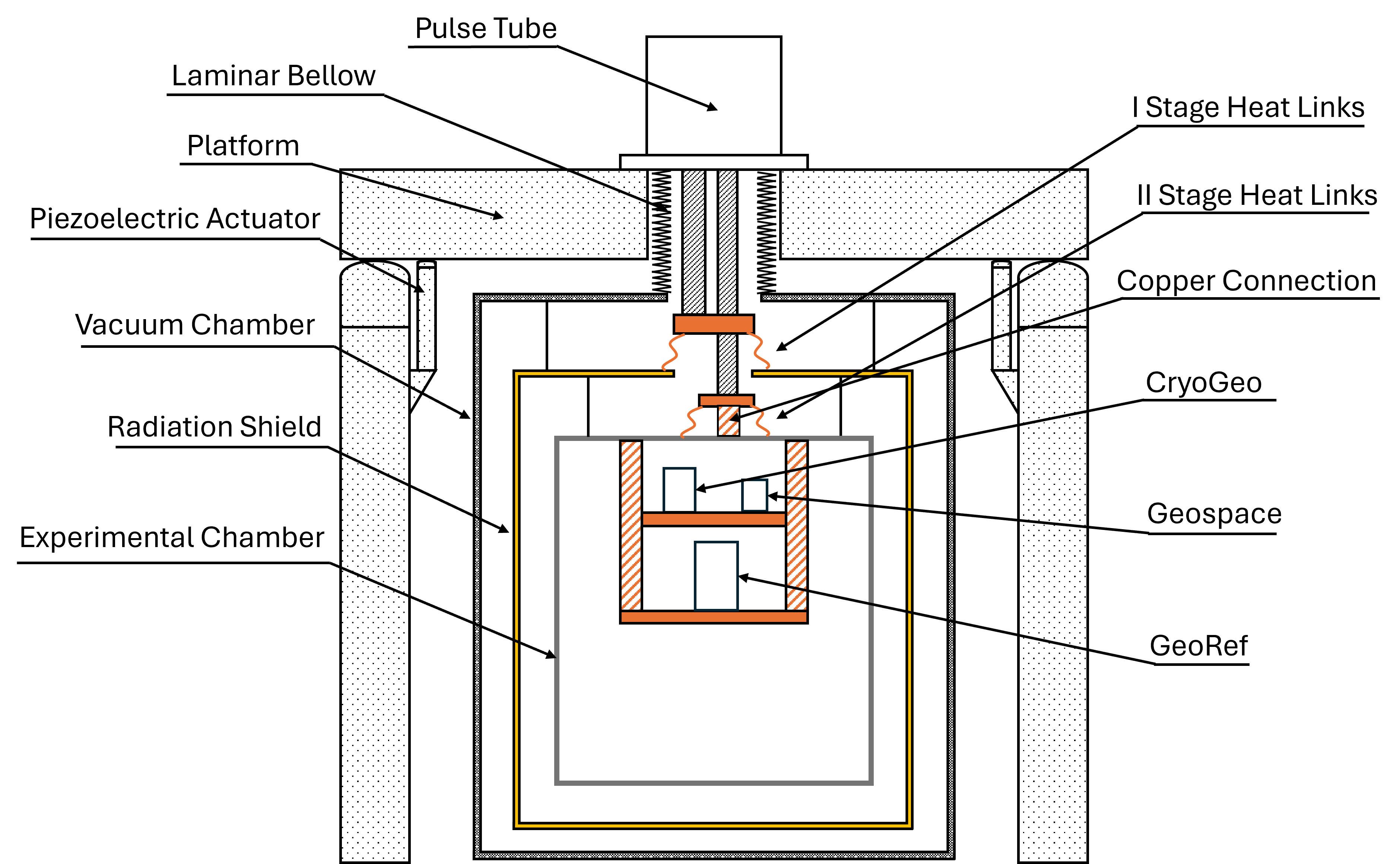}
    \caption{Schematic diagram of the Vibration-Free Cryostat \cite{VFC_Cryostat} configured for dynamic cryogenic calibration. The experimental chamber is rigidly clamped to the PT second stage, bypassing the passive suspension to couple external piezoelectric excitation. Three piezoelectric actuators installed on the support frame drive the entire assembly as an internal shaker.}
    \label{fig:vfc_setup}
\end{figure}

Cryogenic characterization was performed using the Vibration-Free Cryostat (VFC) \cite{VFC_Cryostat, VFC_CRyostatPieo}, schematically illustrated in Fig. \ref{fig:vfc_setup}. The VFC architecture comprises a vacuum vessel housing an intermediate radiation shield wrapped with Multi-Layer Insulation (MLI) and an inner experimental chamber. The PT is supported by a rigid external frame consisting of a top platform and three structural columns (only two are depicted in Fig.~\ref{fig:vfc_setup} for clarity). To decouple environmental vibration, the PT is connected to the vacuum chamber via a flexible laminar metallic bellow. Standard passive vibration isolation is achieved by suspending the radiation shield from the main vacuum chamber, and the experimental chamber from the radiation shield, using stainless steel suspension wires, with thermal anchoring provided by soft copper braid heat links.

However, to enable direct dynamic shaker calibration during cryogenic operation, the passive suspension mechanism was deliberately bypassed. The experimental chamber was mechanically clamped to the second stage of the PT via a rigid copper connection. This intentionally couples ground vibration and external drive signals directly to the experimental space. To emulate the room-temperature electrodynamic shaker setup, three piezoelectric actuators (PI-239-90 \cite{Piezo}) were mounted on the external frame support columns at $120^{\circ}$ intervals.

Inside the experimental chamber, an unmodified reference geophone, \textit{GeoRef} (Seis Tech Inc. \cite{SeisTech}), was installed together with the cryogenic test sensors. To establish an invariant reference baseline during thermal sweeps, \textit{GeoRef} was actively maintained at room temperature ($T = 295\text{~K}$) via a Proportional, Integral, and Derivative (PID) control loop with a dedicated heater and thermometer embedded in its IM. To minimize its thermal impact on the cryogenic environment, the sensor was housed within an auxiliary enclosure wrapped in MLI, suppressing radiative heat transfer. It was also mounted on a rigid, thermally insulating support to limit conductive heat leaks. This thermal decoupling strategy effectively limited the total heating power required to stabilize the reference sensor to approximately $1\text{~W}$ at the lowest base temperature, ensuring that its pre-calibrated room-temperature sensitivity $S_{\text{GeoRef}}(\omega)$ remained perfectly constant throughout the entire cooldown.

Unlike the room-temperature calibration, the cryogenic calibration relies on driving the entire experimental chamber with a frequency sweep applied by the external piezoelectric actuators. Because the test sensors and the thermally stabilized \textit{GeoRef} are rigidly mounted onto the same structure, they are subjected to an identical base displacement $Y(\omega)$. Consequently, the absolute sensitivity of the cooled geophones $S_{\text{cryo}}(\omega, T)$ at any target temperature $T$ can be directly reconstructed by taking the ratio of the measured voltage responses, $V_{\text{cryo}}(\omega, T)$ and $V_{\text{GeoRef}}(\omega)$, scaled by the known room-temperature sensitivity of the reference sensor $S_{\text{ref}}(\omega)$:
\begin{equation}
    S_{\text{cryo}}(\omega, T) = \frac{V_{\text{cryo}}(\omega, T)}{V_{\text{GeoRef}}(\omega)} \cdot S_{\text{ref}}(\omega).
\end{equation}

This approach eliminates the need for an independent, highly complex cryogenic displacement measurement (such as an internal laser interferometer) inside the cryostat.

While this direct piezoelectric excitation approach provides a fundamental validation of cryogenic dynamics, its implementation imposes severe practical limitations. Reconfiguring a standard cryostat to include rigid mechanical bypasses, active thermal management for a reference sensor, and integrated high-load piezoelectric actuators requires extensive hardware modification. This highlights the primary advantage of the step-current method: by requiring only electrical access to the sensor terminals, the step-current approach enables fast, non-invasive, and robust \textit{in-situ} calibration in any standard cryogenic environment without altering the cryostat's mechanical suspension or thermal shielding. 

The direct piezoelectric sweep and the step-current ring-down method are systematically compared across the temperature range for both the unshunted/shunted \textit{CryoGeo} configurations and the commercial \textit{GeoSpace} baseline.
Measurements were recorded at thermal equilibrium across target temperatures ($200\text{~K}$, $130\text{~K}$, $80\text{~K}$, $40\text{~K}$, and $12\text{~K}$), controlled via an active PID loop operating on the experimental chamber. The coil resistance $R_{coil}(T)$ was continuously monitored to track the actual wire temperature (wire RRR $\approx 93$). The thermal characterization of the electromechanical parameters ($G$, $f_0$, and $Q$) across the temperature range validates the design choices implemented in the modified \textit{CryoGeo} sensor compared to the commercial \textit{GeoSpace} unit. Across all test conditions, the results obtained from direct piezoelectric mechanical frequency sweeps align closely with those extracted from non-invasive \textit{in-situ} step-current electrical pulse excitations.

\begin{figure*}[t]
    \centering
    \hspace*{-3cm}
    \begin{subfigure}[b]{0.35\textwidth}
        \centering
        \includegraphics[width=\textwidth, angle=-90]{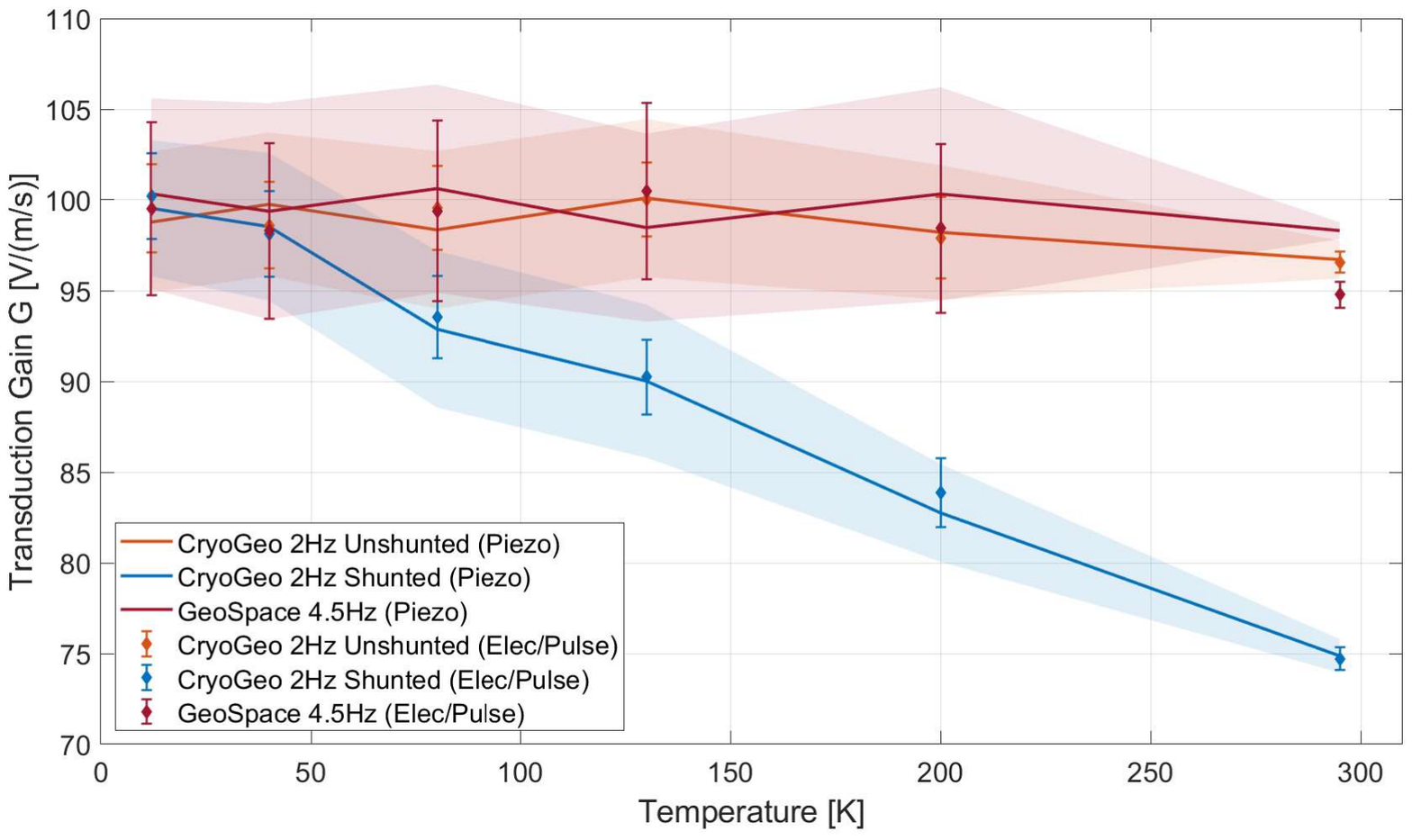}
        \caption{Transduction Gain $G(T)$}
        \label{fig:gain_g}
    \end{subfigure}
    \hspace{0.15\textwidth}
    \begin{subfigure}[b]{0.35\textwidth}
        \centering
        \includegraphics[width=\textwidth, angle=-90]{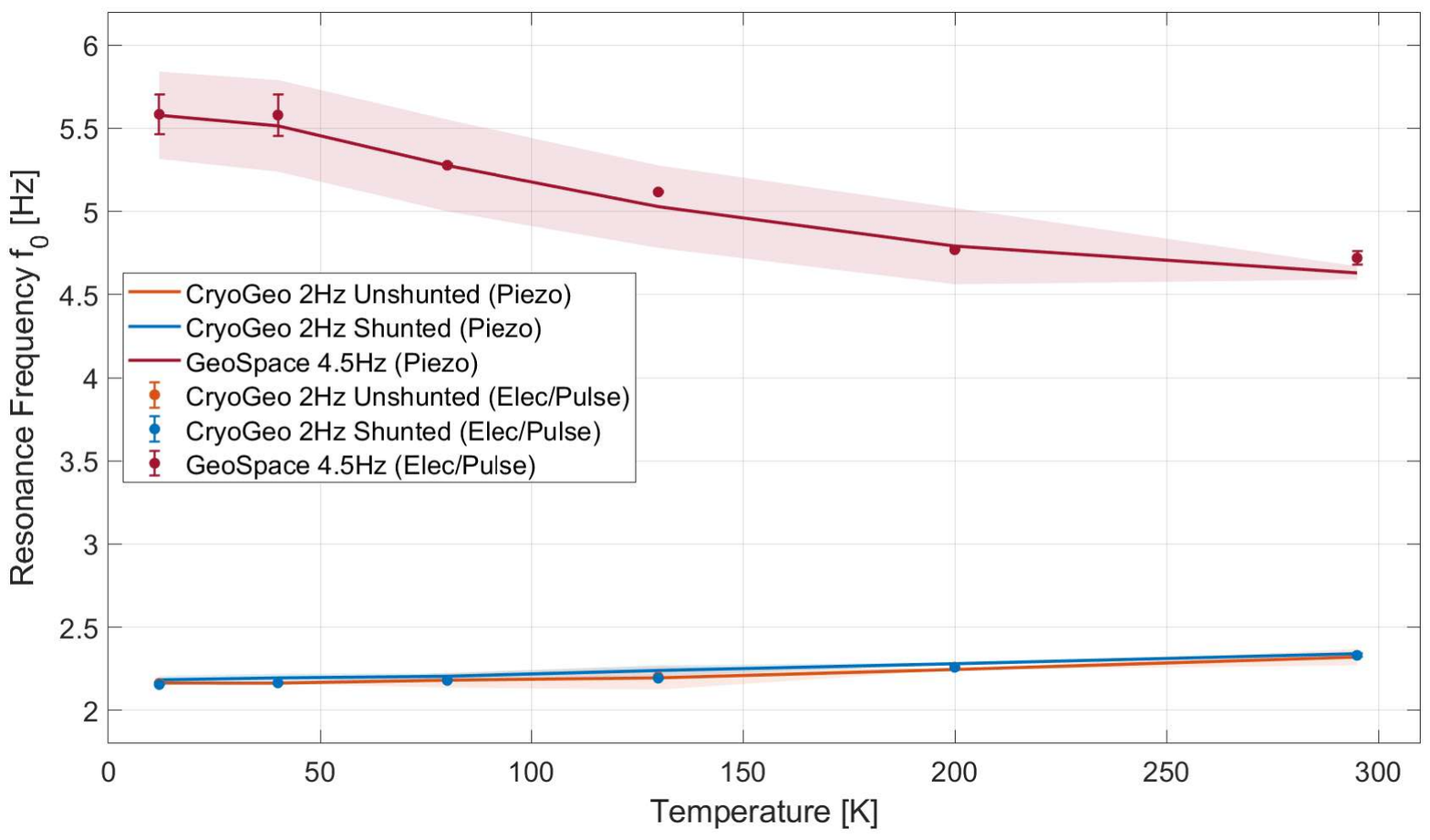}
        \caption{Resonance Frequency $f_0(T)$}
        \label{fig:freq_f0}
    \end{subfigure}

    \vspace{0.01cm} 

    \hspace*{-3cm}
    \begin{subfigure}[b]{0.35\textwidth}
        \centering
        \includegraphics[width=\textwidth, angle=-90]{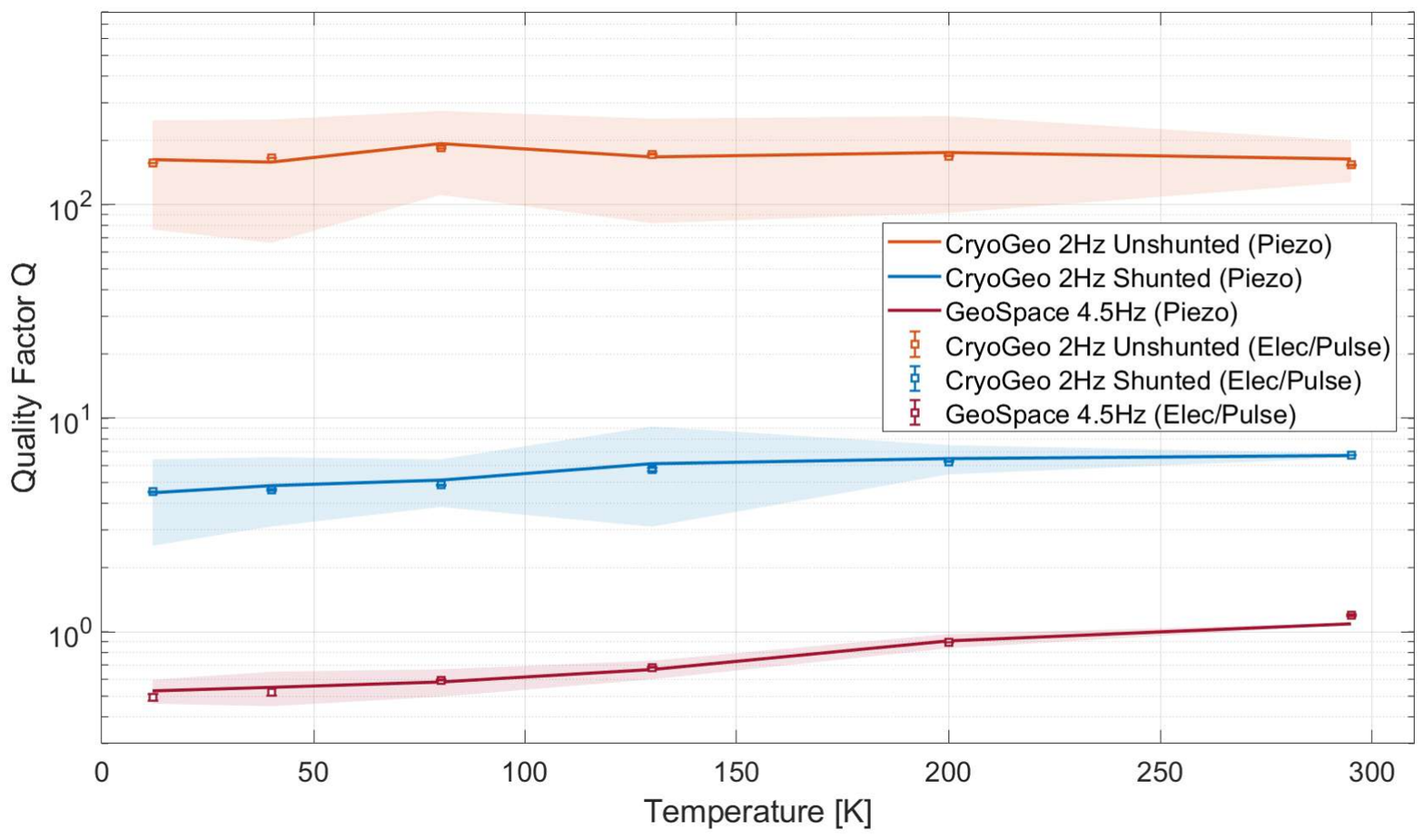}
        \caption{Quality Factor $Q(T)$}
        \label{fig:quality_q}
    \end{subfigure}
    \hspace{0.15\textwidth}
    \begin{subfigure}[b]{0.35\textwidth}
        \centering
        \includegraphics[width=\textwidth, angle=-90]{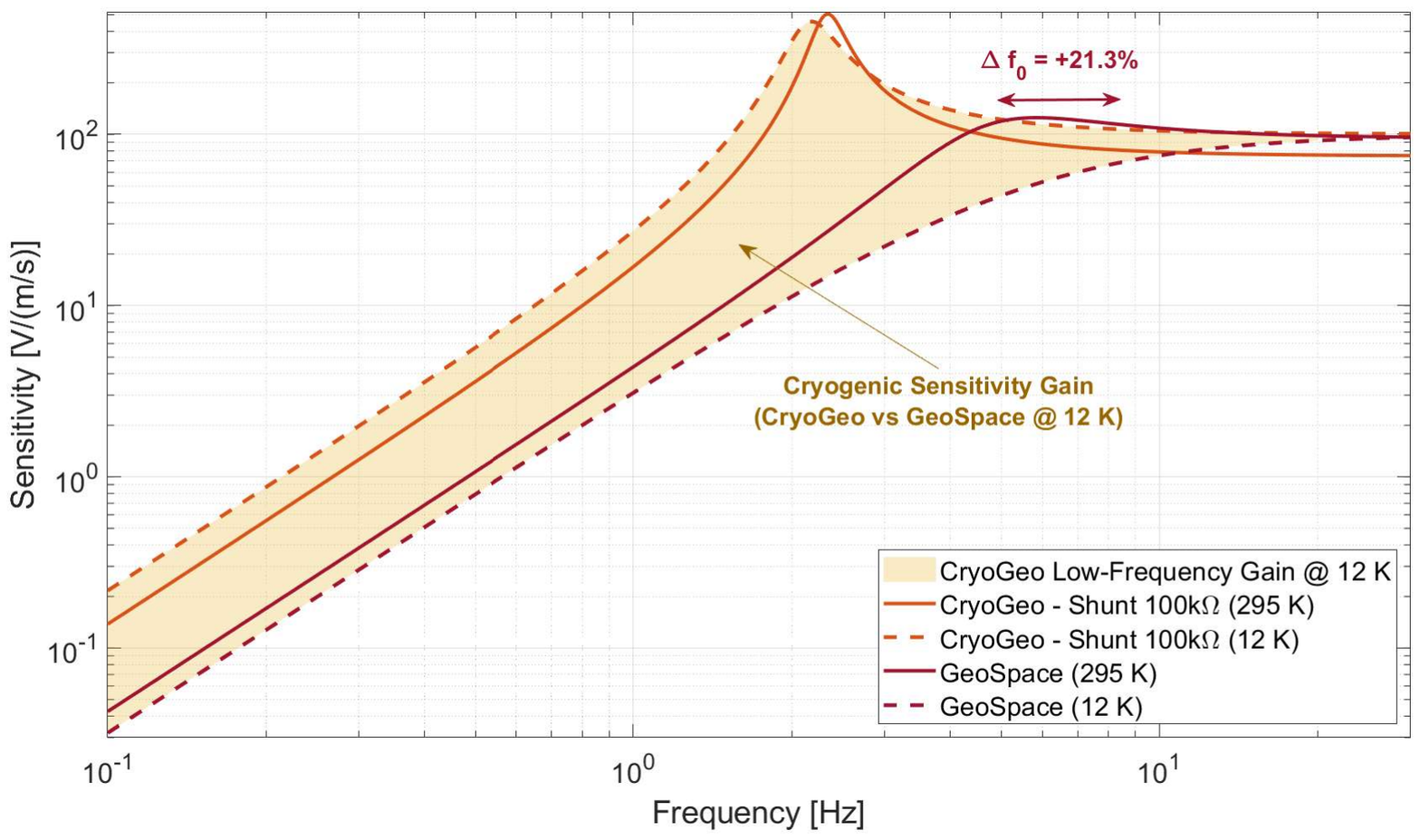} 
        \caption{Cryogenic Sensitivity Advantage}
        \label{fig:sensitivity_gain}
    \end{subfigure}
    
    \caption{Cryogenic thermal performance of the modified \textit{CryoGeo} geophone (shunted and unshunted configuration) compared to the commercial \textit{GeoSpace 4.5Hz} sensor across the $12\text{~K} \le T \le 295\text{~K}$ range: (a) Transduction gain $G(T)$; (b) Natural resonance frequency $f_0(T)$; (c) Quality factor $Q(T)$; (d) Sensitivity comparison $S(\omega)$ highlighting the low-frequency sensitivity advantage of the \textit{CryoGeo} at $12\text{~K}$. Solid lines in (a--c) correspond to piezoelectric frequency sweeps, while markers denote non-invasive electrical step transient measurements.}
    \label{fig:cryo_parameters_combined}
\end{figure*}

\subsubsection{Damping and Quality Factor Dynamics}
The thermal behavior of the quality factor $Q(T)$ (Fig. \ref{fig:quality_q}) highlights the distinct dissipation mechanisms governing the open-circuit, shunted, and commercial configurations:
\begin{itemize}
    \item \textbf{Shunted Configuration:} When the fixed resistive shunt $R_S$ is connected, electromagnetic damping becomes the dominant dissipation channel, and the quality factor follows Eq.~\ref{eq:loaded_parameters}. As the temperature drops to $12\text{~K}$, the copper coil resistance $R_C(T)$ decreases significantly following the RRR. The overall circuit impedance $R_{\text{tot}} = R_C(T) + R_S$ diminishes, enhancing the induced back-electromotive force current and reducing $Q$ from $6.73 \pm 0.01$ at $295\text{~K}$ down to $4.52 \pm 0.05$ at $12\text{~K}$. 
    It should be stressed that $R_S$ was kept constant across temperatures for baseline consistency. However, if a temperature-independent damping factor is required during cooldown, $R_S(T)$ can be actively modulated to compensate for the reduction in $R_C(T)$, thereby maintaining a constant total loop resistance $R_{\text{tot}}$.
    
    \item \textbf{Unshunted Configuration:} For the open-circuit \textit{CryoGeo} sensor, electromagnetic damping is negligible due to the use of an insulating Macor$^{\circledR}$ inertial mass. The quality factor remains consistently high ($Q \approx 150\text{--}185$) across the entire temperature range. The observed minor fluctuations are driven by two main mechanical phenomena: the intrinsic temperature dependence of the spring material's internal friction (characteristic of metallic alloys at cryogenic temperatures \cite{AcousticCopper, DragonCopper}) and the thermal shift of the spring's static equilibrium position.

    \item \textbf{Commercial Sensor Configuration:} In contrast to the custom \textit{CryoGeo}, the commercial \textit{GeoSpace} geophone exhibits a sharp drop in $Q(T)$, falling from $1.09 \pm 0.01$ at room temperature down to a heavily damped regime at $12\text{~K}$ ($Q \approx 0.55 \pm 0.02$). This dissipation is driven by significant parasitic eddy currents induced directly within the conductive metallic body of its inertial mass. (Fig. \ref{fig:Geo_Space_4_5_Hz} shows this effect in the transient pulse response).
\end{itemize}

\subsubsection{Resonance Frequency Stability}
The natural frequency $f_0(T)$ demonstrates the superior mechanical stability of the modified \textit{CryoGeo} design (Fig.~\ref{fig:freq_f0}). While the commercial \textit{GeoSpace} geophone experiences a significant upward drift of over $20\%$ (shifting from $4.60\text{~Hz}$ at $295\text{~K}$ to $5.58\text{~Hz}$ at $12\text{~K}$) due to the thermal stiffening of its standard metallic spring, the \textit{CryoGeo} sensor maintains a remarkably stable resonance at $f_0 \approx 2.33\text{~Hz}$, exhibiting a slight overall shift of less than $3\%$ down to $12\text{~K}$. As expected, the unshunted configuration shows a similar trend for the resonant frequency.
Notably, rather than drifting upward, the resonant frequency of the \textit{CryoGeo} slightly decreases upon cooling. This slight reduction confirms that the thermal increase in Young's modulus $E(T)$ is fully (and marginally over) compensated by the non-linear geometric stiffness reduction as the equilibrium position moves toward the origin $d(T) \to 0$.
Crucially, in the \textit{GeoSpace} unit, this upward shift in $f_0$ acts synergistically with the drastic reduction in $Q$ toward the heavily damped regime ($Q \to 0.5$). The combined effect of a higher resonant frequency and enhanced damping causes severe attenuation of the sensor's low-frequency response below resonance ($f < f_0$). The resulting high-pass roll-off significantly suppresses the measured signal amplitude in the sub-Hz and low-Hz bandwidth, highlighting the superiority of the temperature-compensated design implemented in the \textit{CryoGeo}, see Fig. \ref{fig:sensitivity_gain}.

\subsubsection{Transduction Gain Behavior}
The transduction gain for both unshunted sensors remains constant around $98\text{--}105\text{~V/(m/s)}$, confirming the high magnetic field stability of the internal permanent magnet system down to cryogenic temperatures, see Fig. \ref{fig:gain_g}. 

Conversely, for the shunted \textit{CryoGeo} sensor, the effective transduction gain increases from $74.7 \pm 0.6\text{~V/(m/s)}$ at room temperature to $103.2 \pm 2.0\text{~V/(m/s)}$ at $12\text{~K}$. This trend is fully consistent with the voltage divider effect (see Eq.~\ref{eq:loaded_parameters}) across the shunt: as $R_C(T) \to 0$ ($R_C(T) \ll R_S$), the voltage drop across the internal coil resistance becomes negligible, allowing the measured voltage across $R_S$ to approach the intrinsic open-circuit value. As well as for the quality factor, the external shunt resistance $R_S$ can be actively adjusted during cooldown to dynamically stabilize the effective gain $G'$ at a constant level across all operating temperatures.

\subsection{In-Situ Sub-Kelvin Characterization}

Having validated the equivalence of the step-current method against mechanical shaker sweeps down to $12\text{~K}$, the pulse excitation technique was deployed \textit{in-situ} on a dilution refrigerator platform to extend the electromechanical characterization deep into the sub-Kelvin regime, down to $10\text{~mK}$. 

\begin{figure}
    \centering
    \includegraphics[width=0.42\textwidth]{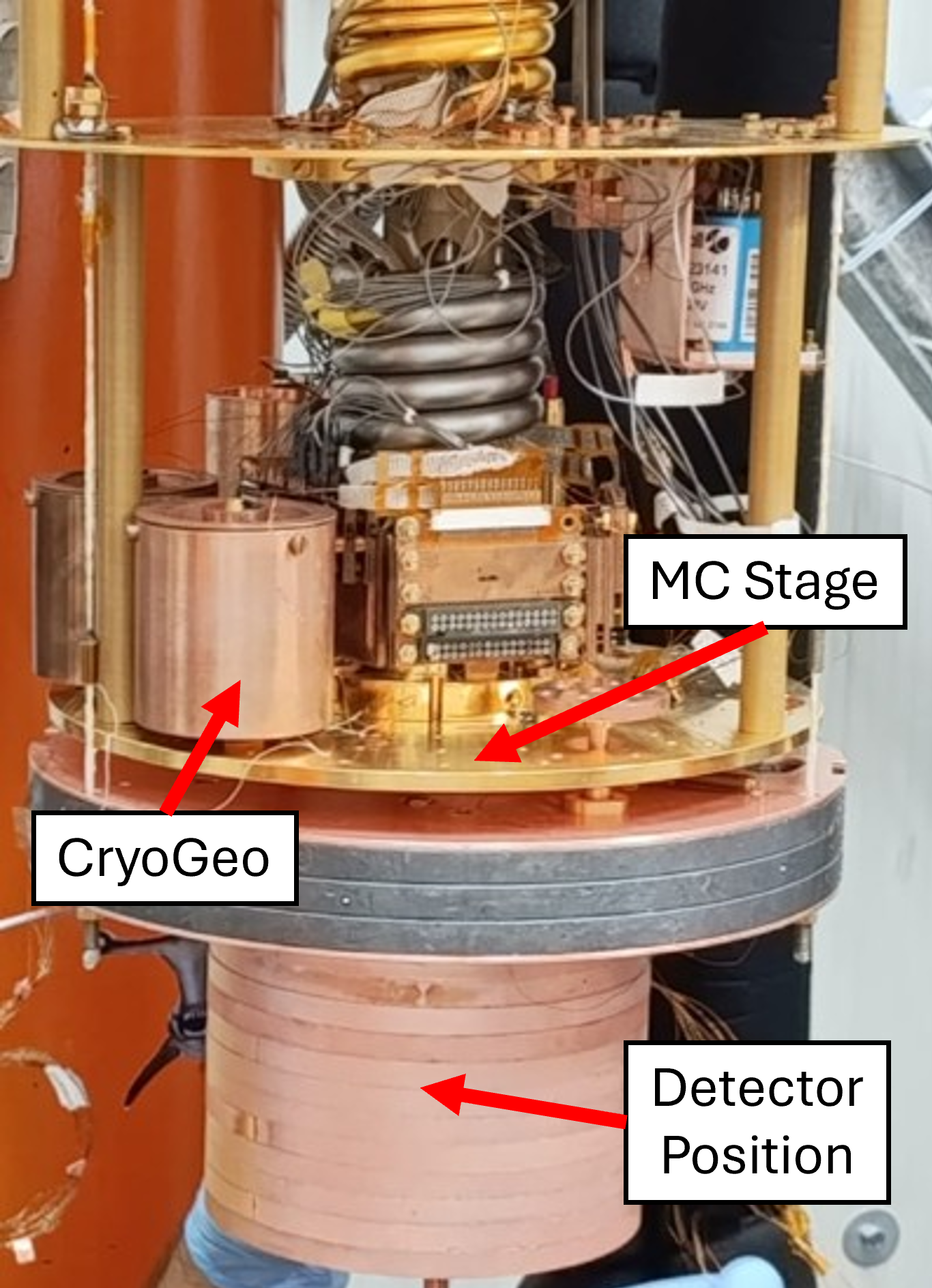}
    \caption{Experimental setup of the modified \textit{CryoGeo} sensor installed at the Ieti cryogenic facility (LNGS). The geophone is rigidly anchored to the MC plate of the dilution refrigerator, ensuring direct conductive thermalization down to the base temperature of $9\text{~mK}$.}
    \label{fig:IETI_setup}
\end{figure}

The experimental measurements were carried out at the \href{https://ieti.sites.lngs.infn.it/}{Ieti} cryogenic facility \cite{IETI_site, IETI} located at the Laboratori Nazionali del Gran Sasso (LNGS). As shown in Fig.~\ref{fig:IETI_setup}, the \textit{CryoGeo} was directly thermally and mechanically anchored to the Mixing Chamber (MC) plate, the coldest stage of the dilution refrigerator, achieving a base operating temperature of approximately $9\text{~mK}$.

The ultra-low heat dissipation inherent to the step-current pulse drive enabled accurate, non-disruptive parameter extraction without compromising the thermal equilibrium at sub-Kelvin temperatures. Specifically, for typical excitation currents, the heat dissipation remains $\mathcal{O}(10\text{~nW})$ over a short calibration pulse duration of $\sim 2\text{~s}$. The extracted sensitivity parameters at $9\text{~mK}$ for both the open-circuit and shunted configurations are summarized in Table~\ref{tab:subkelvin_params}.

\begin{table}
\centering
\caption{Measured sensitivity parameters of the \textit{CryoGeo} sensor at the sub-Kelvin base temperature ($T \approx 9\text{~mK}$) obtained via the \textit{in-situ} step-current method at the Ieti facility (LNGS).}
\label{tab:subkelvin_params}
\begin{tabular}{lccc}
\hline\hline
\textbf{Configuration} & \textbf{$G$ [V/(m/s)]}  & \textbf{$f_0$ [Hz]} & \textbf{$Q$ [-]}\\ \hline
CryoGeo\\ (Unshunted) & $98.6 \pm 0.4$ & $2.27 \pm 0.02$ & $146.8 \pm 1.0$  \\
CryoGeo\\ (Shunted) & $98.5 \pm 0.3$ & $2.27 \pm 0.02$ & $5.23 \pm 0.05$  \\
\hline\hline
\end{tabular}
\end{table}

The parameter comparison confirms that the electromechanical performance saturates below $10\text{~K}$. The parameters remain perfectly preserved down to the sub-Kelvin domain, establishing the suitability of the modified \textit{CryoGeo} design for ultra-low vibration monitoring in mK regime experiments such as rare-event physics searches and quantum computing.

\section{Sensor Self-Noise}
\label{sec:noise}

\begin{figure*}
    \centering
     \includegraphics[width= 0.5\textwidth, angle=-90,clip, trim = 3cm 2cm 3cm 1.4cm]{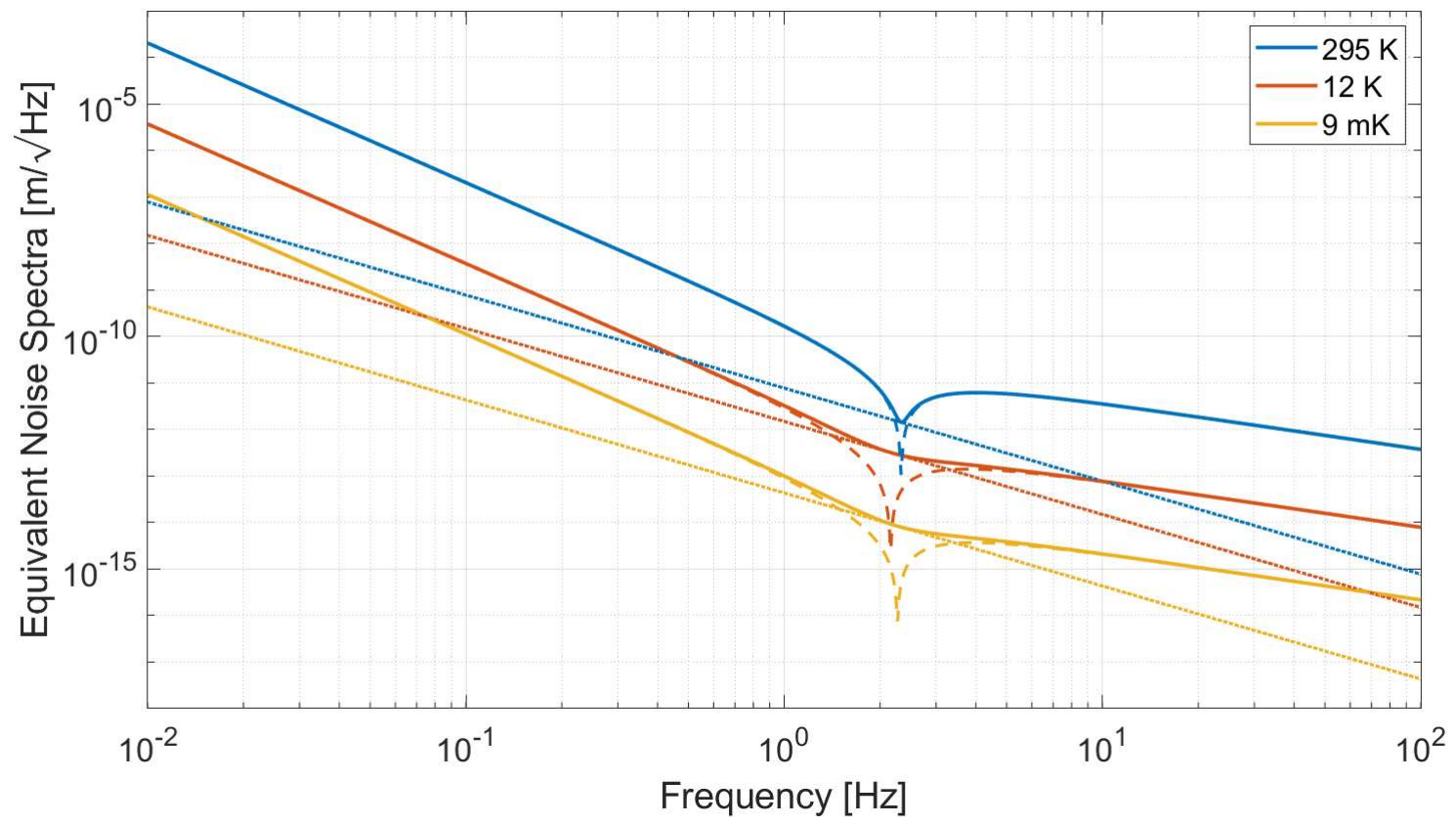}
    \caption{Intrinsic displacement noise power spectra modeled at room temperature ($295$K, blue), $12$K (red), and $9$mK (yellow). For each temperature, solid lines represent the total equivalent self-noise ($\tilde{n}_{TOT}$), dashed lines denote the coil Johnson noise contribution ($\tilde{n}_J$), and dotted lines correspond to the mechanical suspension thermal noise ($\tilde{n}_{th}$). 
    Under cryogenic conditions, the scaling of the noise with both the temperature and the coil's residual electrical resistance (RRR) yields an intrinsic resolution floor down to $10^{-12}\text{--}10^{-13}~\text{m}/\sqrt{\text{Hz}}$ at $1~\text{Hz}$.
    The sharp notch in the Johnson noise reflects the unshunted configuration chosen for the noise estimation.}
    \label{fig:SelfNoise}
\end{figure*}

To evaluate the intrinsic resolution limit of the geophone, its primary self-noise sources were modeled. Neglecting front-end electronic noise, the sensor's resolution is fundamentally constrained by two independent thermodynamic contributions \cite{Barzilai, Kirchoff_HuddleTest}, expressed in terms of equivalent displacement noise spectral density:
\begin{enumerate}
    \item \textbf{Suspension Thermal Noise ($\tilde{n}_{\text{th}}$)}: Originating from the mechanical dissipation within the sensor's suspension system, governed by the Fluctuation-Dissipation Theorem:
    \begin{equation}
        \tilde{n}_{\text{th}}(\omega) = \sqrt{\frac{4 k_B T \omega_0}{m Q}} \times\frac{1}{\omega^2}
    \end{equation}
    where $k_B$ is the Boltzmann constant.
    
    \item \textbf{Coil Johnson Noise ($\tilde{n}_J$)}: Arising from thermal voltage fluctuations across the sensing coil resistance $R(T)$:
    \begin{equation}
        \tilde{n}_J(\omega) = \frac{\sqrt{4 k_B T \, R(T)}}{S(\omega)\>\>\omega}.
    \end{equation}
\end{enumerate}

Assuming uncorrelated noise sources, the total equivalent displacement noise density is obtained by combining the individual contributions in quadrature:
\begin{equation}
    \tilde{n}_{\text{TOT}}(\omega) = \sqrt{\tilde{n}_{\text{th}}^2(\omega) + \tilde{n}_{J}^2(\omega)}.
\end{equation}

As reported in \cite{Kirchoff_HuddleTest}, a geophone's self-noise at room temperature is predominantly dominated by Johnson noise. 
Under cryogenic conditions, both noise contributions decrease significantly, scaling with $\sqrt{T}$. Furthermore, owing to the RRR of the coil wire, the electrical resistance $R(T)$ normally drops by nearly two orders of magnitude. Consequently, the overall theoretical self-noise is reduced by more than an order of magnitude compared to its room-temperature baseline (see Fig.~\ref{fig:SelfNoise} for the noise estimation of the unshunted configuration), yielding an equivalent displacement noise density down to $10^{-13}~\text{m}/\sqrt{\text{Hz}}$ at $1~\text{Hz}$ when operating at $\sim 10~\text{mK}$.

By further optimizing the coil material, using wire with a higher RRR, the Johnson noise contribution could be reduced even further. In this optimized limit, the total self-noise at cryogenic temperatures would eventually be dominated by the mechanical suspension thermal noise ($\tilde{n}_{\text{th}}$), which represents the fundamental limit of the physical sensor.

It should be emphasized that the noise model presented here considers only the intrinsic sensor dynamics. In a practical measurement setup, particularly when probing ultra-quiet seismic environments, the readout signal must be amplified. The noise budget will therefore be augmented by the noise figure of the front-end preamplification electronics, which ultimately limits the overall achievable resolution under experimental conditions.

To mitigate this impact, ongoing developments focus on implementing a dedicated, low-frequency front-end amplifier stage. By adapting low-noise topologies previously developed for cryogenic particle physics detectors \cite{Pessina_Elect}, the target is to achieve an equivalent input voltage noise of a few $\text{nV}/\sqrt{\text{Hz}}$ down to $1~\text{Hz}$. While this electronic noise contribution becomes the dominant factor when operating at ultra-low temperatures ($\sim 10~\text{mK}$), where reaching the sub-$\text{nV}/\sqrt{\text{Hz}}$ regime at $1~\text{Hz}$ remains technologically challenging, it nonetheless preserves a total equivalent displacement noise on the order of $10^{-12}~\text{m}/\sqrt{\text{Hz}}$, effectively maintaining high-sensitivity operation across the entire cryogenic range.

In the low-frequency regime ($\le 1~\text{Hz}$), commercial state-of-the-art cryogenic-compatible accelerometers (such as the Endevco J351B41 \cite{Endevco}) exhibit a room-temperature equivalent displacement noise floor of approximately $2.5 \times 10^{-5}~\text{m}/\sqrt{\text{Hz}}$. 

Even when accounting for the cryogenic thermal suppression of electronic noise, piezoceramic and capacitive transducers remain strongly limited at low frequencies by $1/f$ amplifier noise and charge-readout constraints. In contrast, the proposed cryogenic geophone reaches a projected intrinsic displacement sensitivity of $10^{-12}\text{--}10^{-13}~\text{m}/\sqrt{\text{Hz}}$ at $1~\text{Hz}$. This represents a performance gain of several orders of magnitude, which is crucial for ultra-sensitive low-frequency seismic isolation.

An additional advantage of geophones arises from their passive operating principle. As the sensing element requires neither electrical excitation nor local active electronics, the transduction stage remains electrically isolated from power-distribution networks. This architecture reduces susceptibility to ground-loop currents, mains-frequency interference, and other parasitic noise sources. In addition (in a \textit{CryoGeo}), the comparatively low source impedance, typically a few hundred ohms, results in a robust electrical interface that is significantly less sensitive to capacitive coupling, leakage currents, and external electromagnetic disturbances. These features are particularly beneficial in vibration measurements approaching the intrinsic electronic noise floor.

\section{Conclusions}
\label{sec:conclusions}
In this work, we presented the design, structural optimization, and comprehensive experimental validation of a novel cryogenic geophone capable of delivering a nearly temperature-independent response from room temperature down to ultracryogenic temperatures ($9~\text{mK}$). 

By addressing the fundamental physical mechanisms that induce severe performance degradation in commercial inertial sensors, key structural modifications were implemented:
\begin{itemize}
    \item \textbf{Eddy-Current Suppression:} The standard metallic inertial mass was replaced with non-conductive Macor$^{\circledR}$, completely eliminating parasitic electromagnetic overdamping across the cryogenic range.
    \item \textbf{Elastic and Kinematic Drift Compensation:} A tailored thermal-compensation mass increment ($\Delta m$) was integrated to offset the temperature-dependent increase in Young's modulus of the suspension springs. Combined with expanded internal housing clearances, this strategy dynamically balances non-linear stiffness shifts, maintaining a remarkably stable resonant frequency ($f_0 \approx 2.33~\text{Hz}$, drifting $<3\%$) across the entire $9~\text{mK}$--$295~\text{K}$ span.
    \item \textbf{Conductive Thermalization:} High-purity copper internal supports and chassis provide direct conductive cooling paths, ensuring reliable operation inside dilution refrigerators without introducing thermal leaks.
\end{itemize}

The electromechanical performance was cross-validated using both direct mechanical frequency sweeps and a non-invasive, low-dissipation step-current method. Experimental in-situ characterization on a dilution refrigerator platform confirmed that the sensor parameters fully saturate below $10~\text{K}$ and remain perfectly preserved down to $9~\text{mK}$.

From a noise perspective, the drastic reduction in coil resistance ($RRR \approx 93$) and temperature for cryogenic operation yields an intrinsic equivalent displacement noise floor of $10^{-12}\text{--}10^{-13}~\text{m}/\sqrt{\text{Hz}}$ at $1~\text{Hz}$. This represents a performance gain of several orders of magnitude over state-of-the-art commercial sensors intended for low-temperature use, providing remarkable low-frequency resolution.

Owing to its robustness, absolute parameter predictability, and high operational reliability, this technology has reached Technology Readiness Level (TRL) 8, having been successfully integrated and operated in real cryogenic experimental setups. These achievements have culminated in a joint patent application by INFN and Sapienza University of Rome, opening new avenues for precise low-frequency vibration monitoring and active isolation in rare-event physics experiments, gravitational wave detectors, and quantum computing platforms.

\section*{Acknowledgements}

We express our sincere gratitude to Prof. E. Majorana, to whom we are deeply indebted for his constant support and crucial contributions during the design and testing phases of the cryogenic geophone. The authors warmly thank B. Romualdi for his invaluable expertise and technical skills in the mechanical realization of the first prototype during the early stages of this work. The authors gratefully acknowledge the support of INFN Sezione di Roma, LNGS, and Dipartimento di Fisica at Sapienza Università di Roma for providing the facilities and technical environment that made this work possible.\\
This work was supported by the European Commission (EC) PNRR ETIC project (Grant No. IR0000004 - Missione 4).

\vspace{5mm}
\section*{Conflict of Interest}
The authors declare the following competing interests: S. Pirro, M. Ricci, and S. Cioccolini are inventors on patent priority number IT 102025000014092, owned by INFN and Sapienza Università di Roma, which covers the cryogenic geophone technology described in this work.

\section*{Author Contributions}
\textbf{S. Pirro:} Conceptualization, Data curation, Funding acquisition, Investigation, Methodology, Resources, Supervision, Validation, Visualization, Writing – review \& editing. 
\textbf{M. Ricci:} Conceptualization, Data curation, Formal analysis, Investigation, Methodology, Software, Validation, Visualization, Writing – original draft. 
\textbf{S. Cioccolini:} Methodology, Resources.

\section*{Data Availability}
The data that support the findings of this study are available from the corresponding author upon reasonable request.

\vspace{6pt}


%
%

%


\bibliography{Biblio}

\end{document}